\documentclass[conference]{IEEEtran}
\IEEEoverridecommandlockouts
\usepackage{cite}
\usepackage{amsmath,amssymb,amsfonts}
\usepackage{algorithmic}
\usepackage{graphicx}
\usepackage{textcomp}
\usepackage{xcolor}
\usepackage{url}
\def\BibTeX{{\rm B\kern-.05em{\sc i\kern-.025em b}\kern-.08em
    T\kern-.1667em\lower.7ex\hbox{E}\kern-.125emX}}
\newif\ifmodify 
\modifytrue
\ifmodify

\else

\fi

\usepackage{listings}
\usepackage{color}
\definecolor{bgcolor}{RGB}{255, 255, 214}
\definecolor{commentcolor}{RGB}{76,153,0}
\lstdefinelanguage{P4}
{
    morekeywords={parser, if, control, action, apply, table, key, actions, type, header},
    sensitive=true, % Case sensitive keywords
    morecomment=[l]{//}, % Line comment style
    morecomment=[s]{/*}{*/}, % Block comment style
    morestring=[b]", % String literal style
    morestring=[b]', % Character literal style
}

\begin{document}
\title{Extracting TCPIP Headers at High Speed for the Anonymized Network Traffic Graph Challenge
\thanks{This  work  was  funded  by  National  Science  Foundation (NSF) grants CNS-1925464, CNS-1925658, CNS-2130891, and 2130907. All opinions and statements in the above publication are of the authors and do not represent NSF positions.}}
\author{\IEEEauthorblockN{Zhaoyang Han\IEEEauthorrefmark{1},
Andrew Briasco-Stewart\IEEEauthorrefmark{3}, 
Michael Zink\IEEEauthorrefmark{2}, Miriam Leeser\IEEEauthorrefmark{1}}
\IEEEauthorblockA{Northeastern University \IEEEauthorrefmark{1}\IEEEauthorrefmark{3}, University of Massachusetts Amherst \IEEEauthorrefmark{2}\\
Email: \IEEEauthorrefmark{1}{zhhan,mel}@coe.neu.edu, \IEEEauthorrefmark{3}briasco-stewart.a@northeastern.edu \\
\IEEEauthorrefmark{2}mzink@umass.edu,}
}

\maketitle

\begin{abstract}

Field Programmable Gate Arrays (FPGAs) play a significant role in computationally intensive network processing due to their flexibility and efficiency. Particularly with the high-level abstraction of the  P4 network programming model, FPGA shows a powerful potential for packet processing. By supporting the P4 language with FPGA processing, network researchers can create customized FPGA-based network functions and execute network tasks on accelerators directly connected to the network. A feature  of the P4 language is that it is stateless; however,  the FPGA implementation in this research requires state information.  This is accomplished using P4 externs to describe the stateful portions of the design and to implement them on the FPGA using High-Level Synthesis (HLS).   This paper demonstrates using an FPGA-based SmartNIC to efficiently extract source-destination IP address information from network packets and construct anonymized network traffic matrices for further analysis.  The implementation is the first example of the combination of using P4 and HLS in developing network functions on the latest AMD FPGAs.  Our design achieves a processing rate of approximately 95 Gbps with the combined use of P4 and High-level Synthesis and is able to keep up with 100 Gbps traffic received directly from the network.

\end{abstract}


\section{Introduction}
%%1. Describe Large Scale Network problems
%%2. How Graph Challenge can help with these problems
%%3. Describe FPGA potential in computational intensive networking processing

Large-scale network problems represent some of the most pressing challenges in our increasingly connected world.  These problems encompass a variety of complex issues related to the design, analysis, and optimization of vast networks that facilitate global communication and data exchange. As networks continue to grow in size and complexity, addressing these problems becomes crucial for ensuring efficient, reliable, and secure operations.  A significant bottleneck in the current network infrastructure is its inability to process large network traffic while maintaining the flexibility needed to adapt to varying demands and conditions.

The MIT Graph Challenge~\cite{graphChallenge} is an important platform for tackling large-scale network problems. By providing a structured competition focused on innovative graph analytics, the challenge encourages researchers and practitioners to develop cutting-edge solutions.  Participants engage with real-world datasets and complex graph problems, driving advances in algorithms and technologies that can be applied to optimize large-scale networks. In particular, network researchers use a data product of anonymized source-to-destination traffic matrices derived from billions of real network packets to analyze and solve real large-scale network problems.

The Anonymized Network Sensing Graph Challenge~\cite{Jeremy2024Hpec} is part of the Graph Challenge that aims to
%to facilitate large-scale, open, and community-driven methods for network problems. Especially, the goal of the Anonymized Network Sensing Graph Challenge is to 
find optimized and highly efficient approaches in the construction of anonymized traffic matrices from network traffic.  
As part of the Anonymized Network Sensing Graph Challenge, we proposed a novel approach based on the  Programming Protocol-independent Packet Processors (P4) language combined with the implementation on Field Programmable Gate Arrays (FPGAs). 

FPGAs have demonstrated significant potential in computationally intensive network processing, especially when they are directly connected to the network~\cite{leeser2021fpgas}.  Their ability to offload compute-intensive tasks and support the disaggregation of data centers makes them valuable in addressing the demands of modern network infrastructures. FPGAs can be directly connected to networks, enhancing their role in processing and managing large volumes of data efficiently. In addition, the reconfigurability of FPGAs allows them to be programmed and reprogrammed to perform specific tasks.  This flexibility is particularly beneficial for network processing, where requirements can change rapidly. FPGA-based network devices can be tailored to accelerate various networking functions, such as packet filtering, encryption, and deep packet inspection, and can be updated as new protocols and standards emerge.
The P4 language is a high-level abstraction for network devices that allows the programmer to describe how network packets should be parsed, processed, and forwarded. The combination of the P4 language and FPGA implementation allows network researchers to easily develop their own FPGA-based network functions~\cite{han2023framework,balp4}.
 
In this research we focus on constructing the anonymized data structure of network traffic as described in the Anonymized Network Sensing Graph Challenge.  This innovative solution makes use of FPGAs programmed with P4 to form a powerful approach to match the line rate of the high-speed network. Specifically, we demonstrate the use of an FPGA-based SmartNIC that is able to perform tasks for the Anonymized Network Sensing Graph Challenge at approximately 95Gbps rate. 

The contributions of the work presented in this paper are to:
\begin{itemize}
    \item Describe the design flows of FPGA-based SmartNICs in developing high-throughput network functions using P4. In particular, this paper shows the first example of the combination of using P4 and High-level Synthesis in developing network functions on the latest AMD FPGAs. 
    \item Demonstrate innovative FPGA-based SmartNIC solutions for the Anonymized Network Sensing Traffic Graph Challenge.
    \item Implement and test the solution with the CAIDA dataset on the public Open Cloud Testbed (OCT), achieving high-throughput processing at 95Gbps.
\end{itemize}

The rest of this paper is organized as follows. In Sec.~\ref{sec:background} we introduce the Anonymized Network Sensing Traffic Graph Challenge problem, discuss using the P4 language for programming FPGAs, and describe the Open Cloud Testbed (OCT) that is used for these experiments.  Our design is described in Sec.~\ref{sec:design} and results are presented is Sec.~\ref{sec:eval}.  
We present a discussion of future directions for incorporating this design into  the rest of the graph challenge problem in Sec.~\ref{sec:discuss} and conclude the paper with a summary of our design, evaluation and future directions.

\section{Background}
\label{sec:background}

In this section, we introduce the details of the  Anonymized Network Traffic Graph Challenge problem, the FPGA-based P4 SmartNICs used in our implementation, the Open Cloud Testbed (OCT) as well as discussing related research.

\subsection{Anonymized Network Sensing Challenge}

\begin{figure}[ht]
  \center
  \includegraphics[scale=1.3,width=\linewidth]{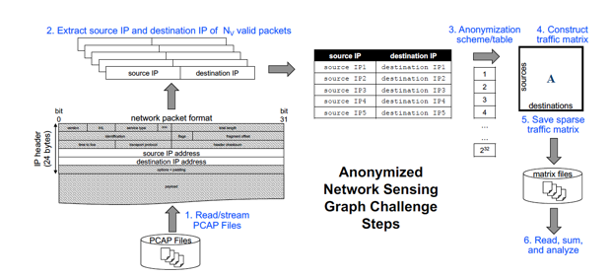}
  \caption{\label{fig:steps}Anonymized Network Traffic Graph Challenge Steps. Figure adapted from \cite{Jeremy2024Hpec}.}
  \vspace{-10pt}
\end{figure}

The Anonymized Network Traffic Graph Challenge aims to extract source and destination information from real network traffic, as shown in Fig.~\ref{fig:steps}. Our solution focuses on the first two steps. To simulate real network traffic, participants are encouraged to use network traces captured in real networks for the evaluation. 
%Each step of the process can be improved and optimized. 
To test the performance of the flow, we use the realistic network traffic from the Center for Applied Internet Data Analysis (CAIDA)~\cite{CAIDA}.

\subsection{P4}

The P4 language is a domain-specific language to describe how network devices like switches, Network Interface Cards (NICs), routers, filters, etc.\ should process packets. This device-agnostic high-level abstraction provides network researchers an opportunity to design their network functions without considering hardware devices.

The evolution of the P4 language~\cite{bosshart2014p4} and the advent of programmable network devices have significantly enhanced the user's ability to program both control and data planes. This progress has enabled extensive research in areas such as data plane disaggregation, cryptography, and data plane machine learning.  Several P4 programmable devices are available including AMD Pensando, Intel Tofino Switch, NVIDIA Bluefield2 SmartNICs,  etc.
FPGAs are a key technology to empower the performance of packet processing in the data plane~\cite{sultana2021flightplan, yazdinejad2020p4, swamy2022taurus}, and can be programmed with the P4 language, as described in the next section.  

\subsection{FPGAs as Network Devices}

There is a trend to attach more devices directly to the network to improve data access and decrease latency.  Modern applications, including machine learning~\cite{diaconu2023machine}, require a large amount of data that can be more efficiently accessed directly from the network. SmartNICs are gaining popularity as a packet processing platform and can be programmed using the P4 programming model.

FPGA-based SmartNICs offer superior programmability and customization compared to dedicated P4 SmartNICs.  Compared to AMD's Pensando Distributed Services Card (DSC-200), which has a fixed data flow and lacks flexibility,  SmartNICs based around FPGA programming can be customized and scaled to address specific requirements. Firestone et al.~\cite{firestone2018azure} demonstrated this advantage by developing and deploying FPGA-based SmartNICs on Azure, effectively offloading network functions.  Their results show that FPGAs excel at dynamically managing network traffic, alleviating the main CPU processing burden.
We have created a framework for developing and testing network functions on P4-based FPGA SmartNICs using the Open Cloud Testbed, a public research platform described in Sec.~\ref{sec:oct}. 
This P4-based framework offers a more efficient method for describing network behaviors on FPGAs than traditional Hardware Description Languages (HDLs) such as Verilog and VHDL.  While FPGAs are ideal for developing P4 applications due to their customizable pipelines, the P4 model does not fully leverage the advantages of FPGA parallel programming. Our framework extends the original P4 model with the extra ability for concurrent processing using the \texttt{extern} function, a P4 language construct. Such extern functions can be described in HDL or C++ and translated to the FPGA fabric using either a standard tool flow or  High-Level Synthesis (HLS).  We utilize the P4 framework in OCT to develop our solution for the Graph Challenge due to its ease of use and high performance.

\subsection{Open Cloud Testbed}
\label{sec:oct}

\begin{figure}[ht]
    \center
    \includegraphics[scale=1.3,width=\linewidth]{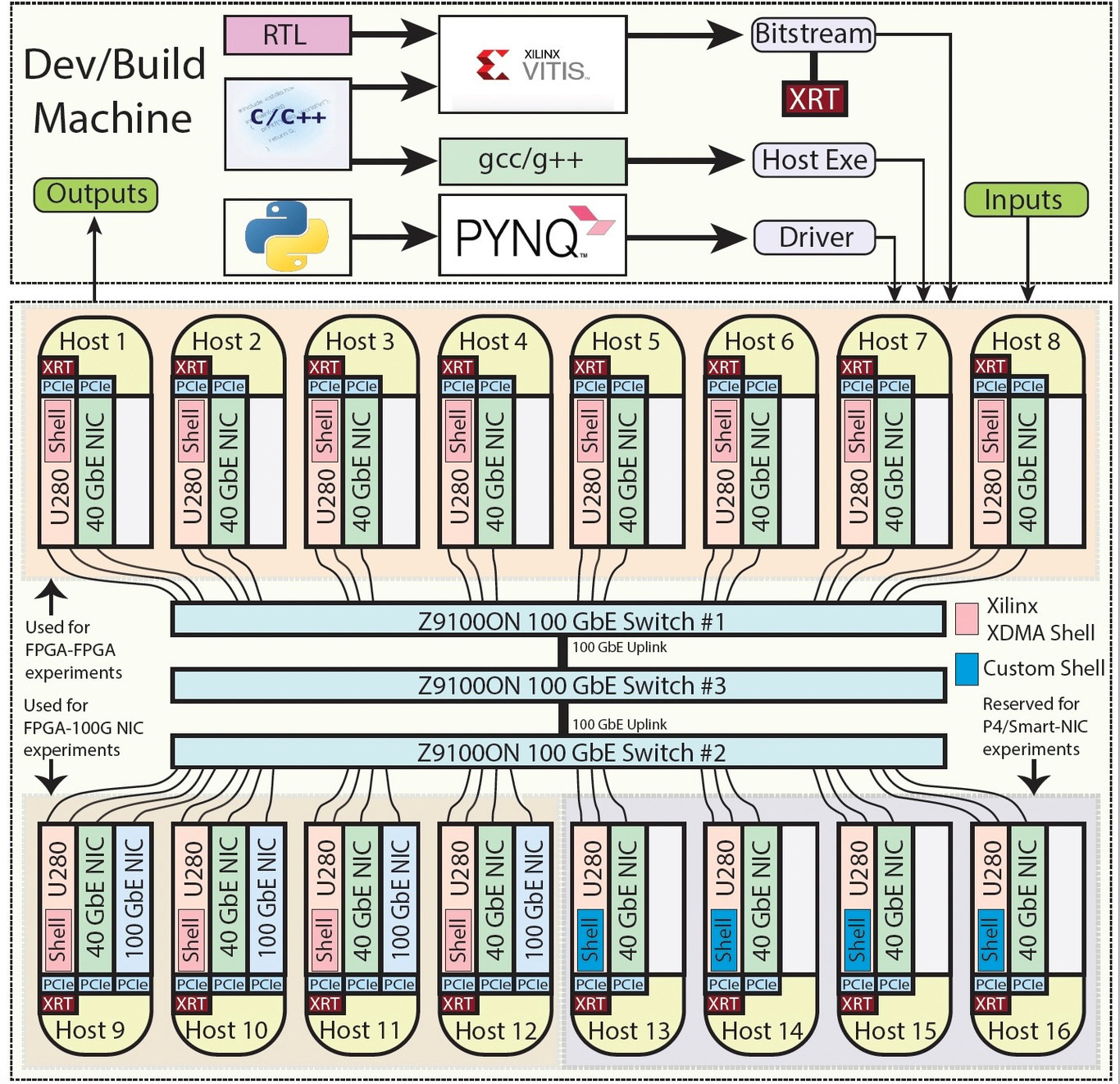}
    \caption{\label{fig:oct-setup}Overview of OCT FPGA development and workflow\cite{zink2021open}}
    \vspace{-10pt}
  \end{figure}
  
The Open Cloud Testbed (OCT)~\cite{zink2021open} provides a research-oriented experimentation testbed for systems researchers who focus on cloud platforms. Testbeds like OCT deliver the necessary hardware and software on top of bare metal services to researchers in both the cloud and system communities, enabling more experimental-based research.

OCT currently offers 32 FPGAs to the research community: 24 AMD Xilinx Alveo U280, 4 AMD Xilinx VCK5000, and 4 AMD Xilinx V70. Each of these FPGAs is housed in a host server that an experimenter can allocate as a bare metal machine. 28 (all U280s and VCK5000s) of the 32 FPGAs have two direct network links, which connect both of their QSFP28 (100GbE) interfaces to a switch for a combined maximum bandwidth of 200 Gbps~\cite{handagala2022network}. In tandem, OCT offers a toolchain that supports the development of bitstreams that can be deployed on the FPGAs~\cite{leeser2021fpgas}.

% with High Bandwidth Memory to support data-heavy tasks

The OCT workflow consists of two primary stages which are illustrated in Fig.~\ref{fig:oct-setup}.  We provide a series of tutorials, example applications, and scripts and profiles for the setup and execution of experiments~\cite{oct_git}.

\textbf {Development Stage:} OCT development tools are hosted on a virtual machine (VM) within the New England Research Cloud (NERC). OCT users can remotely log into this VM to create FPGA bitstreams, host executables, and drivers using the provided tools. In addition, licenses required for certain Xilinx IPs are hosted on a separate license server.
OCT provides several different FPGA configurations that can be used for different research directions and are well-suited to this research.

This basic framework is available on OCT as the HLS acceleration flow. Users can use HLS to develop their compute-intensive accelerators and test them on OCT. Recently, P4-based FPGA support has been made available on OCT, including tutorials and several demonstration examples~\cite{han2023framework}. The P4 design further supports the potential of FPGAs as programmable network devices.   

In this paper, we make use of the P4-based tool flow as the basic framework and combine it with HLS. Specifically, this paper demonstrates the first example of the combination of P4+HLS on the FPGA-based SmartNICs.

\textbf{ Deployment Stage:} After creating the bitstreams and host executables/drivers, users transfer them to bare metal servers that host the FPGAs. The subsequent process involves programming the FPGAs, executing the host executables, and optionally fetching the results back to the development machine. 

% OLD 

% The Open Cloud Testbed~\cite{zink2021open} offers a research-focused experimental platform for systems researchers specializing in cloud technologies.  OCT currently includes 237 servers with a total of 5,172 cores and 63TB of RAM, as well as 32 FPGAs.  The FPGAs are AMD/Xilinx Alveo U280 data center cards, each  equipped with 2 100Gbps network interfaces. These FPGAs are connected through a top-of-rack switch as shown in Fig.~\ref{fig:oct-setup}.  In 2021, the initial deployment of 16 FPGAs, along with the corresponding toolchain for acceleration tasks, was completed on OCT~\cite{leeser2021fpgas}.  In 2022, support for direct network connections from the FPGAs was added~\cite{handagala2022network}, supporting the FPGAs as network-attached accelerators.  In 2023, P4-based FPGA support became available on OCT, including tutorials and several demonstration examples~\cite{han2023framework}. The P4 design further supports the potential of FPGAs as programmable network devices.  In summary, OCT provides several different FPGA configurations that can be used for different research directions.

\subsection{Related Work}

Several studies have explored the potential of FPGAs in-network processing due to their customizable pipelines and high throughput. Since the inception of P4~\cite{ibanez2019p4,bosshart2014p4}, various compilers and frameworks have been developed for FPGAs. In 2023, ESNet introduced a framework that integrates the latest AMD FPGA with their VitisNetP4 compiler~\cite{esnet}. Similarly, \cite{han2023framework} presented a framework that offers additional features such as support for \texttt{extern} functions and partial reconfiguration. In this work, we extend the framework from \cite{han2023framework} by utilizing simplified C-based HLS functions as extern functions.
For the graph challenge, others have investigated the performance of deploying a solution on the edge with two Accolade Technology ANIC-200Kq dual port 100 gigabit NICs \cite{9926332}. Their work demonstrates an example of the anonymized networking sensing challenge with high throughput. Our design achieves a comparable throughput as ~\cite{9926332}.  Similar to this work, we deploy our solution on an FPGA-based NIC. Our work demonstrates comparative results that saturate the 100Gbps link on the FPGA and also showcases a development with state-of-the-art domain-specific language P4.

\section{Proposed Design}
\label{sec:design}

\begin{figure}[ht]
    \center
    \includegraphics[scale=1.3,width=\linewidth]{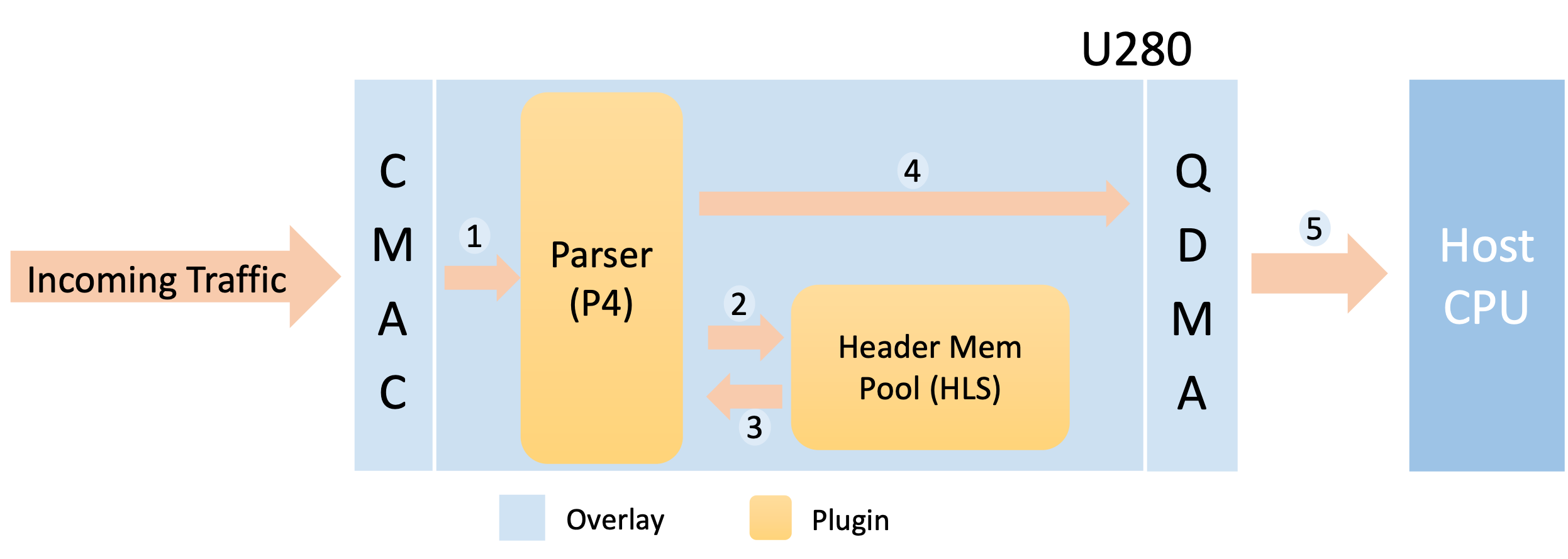}
    \caption{\label{fig:design}Design illustration: the numbers shows the order of data flows.}
    \vspace{-10pt}
  \end{figure}

We proposed a design solution for the Anonymized Network Sensing Challenge that focuses on the first two steps shown in Fig.~\ref{fig:steps}. This optimized hardware design provides a high-performance solution to in-network packet processing and header extraction.
In this section, we explore the utilization of our P4-based FPGA framework for implementing high-performance packet header extraction.  The design consists of an overlay and a plug in.  The overlay establishes a shell that facilitates basic connections between physical interfaces and host CPUs, and is common among P4 designs.   The plugin houses the P4 application including the extern.  The P4 parts handles the packet parsing and deparsing and the HLS parts are used to save the extracted headers..  The framework consisting of the the overlay and  plugin is shown in Fig.~\ref{fig:design}. 

\subsection{Overlay Design}

The overlay is built on AMD/Xilinx's OpenNIC shell~\cite{opennic}, an open-source, FPGA-based 100G NIC platform. The OpenNIC shell features a Queue-based Direct Memory Access (QDMA) block for transferring packets between the NIC shell and the host CPU via the PCIe bus, and a 100G MAC block (CMAC) for Ethernet communication. The user plugin connects the QDMA and CMAC components.

\begin{figure}[ht]
    \center
    \includegraphics[scale=1.3,width=\linewidth]{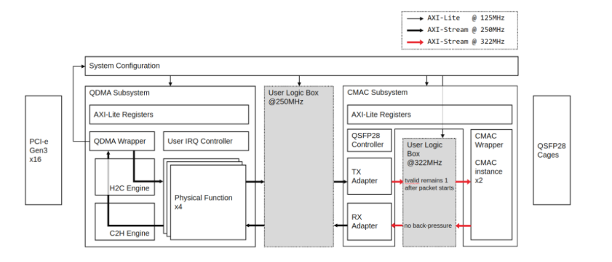}
    \caption{\label{fig:nic}OpenNIC shell structure. Figure adapted from the OpenNIC shell documentation~\cite{opennic}}
    \vspace{-10pt}
  \end{figure}
  
As shown in Figure~\ref{fig:nic}, we populate the block labeled \emph{User Plugin@250MHz} with the P4 hardware IP block produced by AMD/Xilinx's Vitis Networking P4 (VitisNetP4) toolchain~\cite{vitisnet}. This toolchain generates hardware IP from P4 source code. The OpenNIC shell delivers NIC functionality capable of supporting 100Gb/s throughput. Our framework establishes the required packet and control logic pathways between the OpenNIC shell and the outputs from VitisNetP4.

With this approach, users only needs to provide P4 code as input, and can quickly generate a P4-enabled NIC and achieve fast deployment on OCT using this tool flow. It provides an easy way for network researchers unfamiliar with FPGAs or hardware design to conduct their research on OCT. 
Under this framework, users can easily focus on network functions while the OpenNIC shell provides  the development of the NIC design.

\subsection{Header Extraction}

The key to achieving the challenge lies in utilizing the plugin.  This plugin is a combination of P4 and HLS blocks. The P4 block consists of a parser and a deparser. The parser is employed to extract source and destination IP addresses, while the deparser is used to reconstruct the packets and forward them to the host.  A limitation of the P4 processing model is its stateless nature; the extracted header information cannot be stored under the P4 description.  For this reason, we use an HLS block as a P4 \texttt{extern struct} to describe the behaviors of storing and processing this information on the FPGA. The HLS-based extern function is instantiated in the processing block between the parser and deparser of the P4 code. 

We store the extracted headers on the FPGA on-chip Block RAM. Since the traffic matrix is constructed on the host CPU, we pack the information from a fixed number of $N_p$ packets received from the network into one packet and forward it to the host CPU. Similar to a network packet,  we need eight bytes for the address of these information packets. Therefore, a fixed size $8\times N_p$ bytes network packet is sent to the host. It will yield a theoretical packet drop rate $1\div (N_p+1)$. Although the larger $N_p$ yields a lower drop rate, it also requires a wider data bus between the P4 block and HLS block to exchange information as $8\times N_p$ data need to be transferred. A wide data bus will increase the hardware design difficulties and eventually cause a larger packet drop rate than expected. This design accommodates the current FPGA configuration. To further improve the performance and reduce the drop rate, instead of sending data as a network packet, we can establish a direct data link between the HLS block and the host as described in section~\ref{sec:discuss}.

\section{Implementation and Evaluations}
\label{sec:eval}

\begin{figure}[ht]
    \center
    \includegraphics[scale=1.3,width=\linewidth]{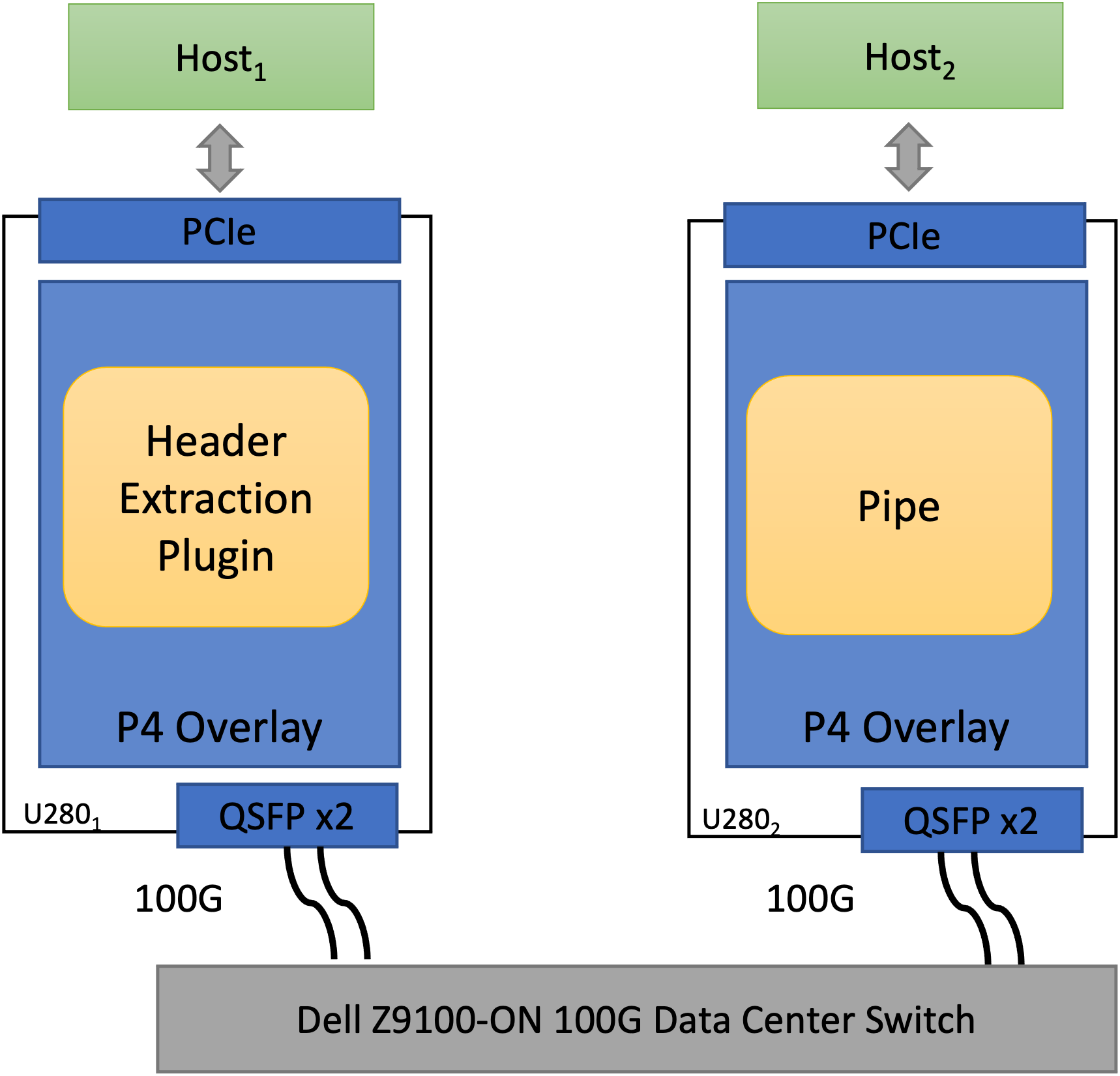}
    \caption{\label{fig:testbed}Testbed Illustration}
    \vspace{-10pt}
  \end{figure}

To evaluate the performance of our solution, we implemented it on OCT, as illustrated in Fig.~\ref{fig:testbed}. We deployed our solution on one FPGA node in the OCT as the receiver. To generate a sufficient number of packets for testing, we utilized another U280 FPGA node as the sender. Both nodes are capable of handling 100Gbps traffic through their network connections. On the sender side, with an empty plugin, packets are directly transmitted through the FPGA. To fully utilize the hardware's speed and minimize performance bottlenecks from the host CPU processing rate, we also employed DPDK on both hosts. 

\begin{figure}[htb]
\begin{lstlisting}[caption={P4-based front-end}, label=lst:p4_codes,xleftmargin=2.5ex]
...
parser MyParser(packet_in packet,
                out headers hdr,
                inout metadata meta,
                inout standard_metadata_t smeta) {
    state start {
        transition parse_eth;}
    state parse_eth {
        packet.extract(hdr.eth);
        transition select(hdr.eth.type) {
            IPV4_TYPE : parse_ipv4;
            default   : accept;}}
    state parse_ipv4 {
        packet.extract(hdr.ipv4);
        packet.extract(hdr.ipv4opt, (((bit<32>)hdr.ipv4.hdr_len - 5) * 32));
        transition select(hdr.ipv4.protocol) {
            TCP_PROT  : parse_tcp;
            UDP_PROT  : parse_udp;
            default   : accept;}}
    state parse_tcp {
        packet.extract(hdr.tcp);
        packet.extract(hdr.tcpopt, (((bit<32>)hdr.tcp.dataOffset - 5) * 32));
        transition accept;}
    state parse_udp {
        packet.extract(hdr.udp);
        transition accept;}
}
...
\end{lstlisting}
\end{figure}
\begin{table}[ht]
\caption{\label{table:results}FPGA Processing Rate}
\begin{tabular}{|l|l|l|}
\hline
Packet Size (Byte) & Data Rate (Mbps) & Packet Rate (pps) \\ \hline
64                 & 27,704           & 41,210,656        \\ \hline
128                & 47,432           & 39,790,209        \\ \hline
256                & 77,718           & 35,098,591        \\ \hline
512                & 94,238           & 22,428,831        \\ \hline
1024               & 95,759           & 11,447,680        \\ \hline
1518               & 95,359           & 7,746,182         \\ \hline
\end{tabular}
\end{table}

Table~\ref{table:results} shows the performance results of processing different size packets using our implementation. We use the CAIDA Anonymized Internet Traces Dataset~\cite{CAIDA}.  In particular, we use the traces captured by the Equinix-Chicago monitor on high-speed internet backbone links in 2016. These traces only contain the TCP/IP headers. To test on a real network, we pad these packets with zeros to simulate real traffic data. Table~\ref{table:results} shows that for 512-byte packets, we can fully saturate the 100Gbps link. The packet rate for small packets can be further improved as the theoretical performance of the OpenNIC shell overlay is 100 million packets per second.

Through the use of P4 and HLS, we were able to reduce the effort of developing hardware logic. The FPGA design consists of a few hundred lines of P4 codes in place of thousands of lines of HDL code.  The shorter code base makes it easier to develop, debug, and improve our FPGA design. Listing.~\ref{lst:p4_codes} shows the P4 parser code to extract the headers.  If this code were developed in HLS, thousands of lines of detailed logic description would have been required to describe the parsing logic.

\begin{figure}[ht]
    \center
    \includegraphics[scale=1.3,width=\linewidth]{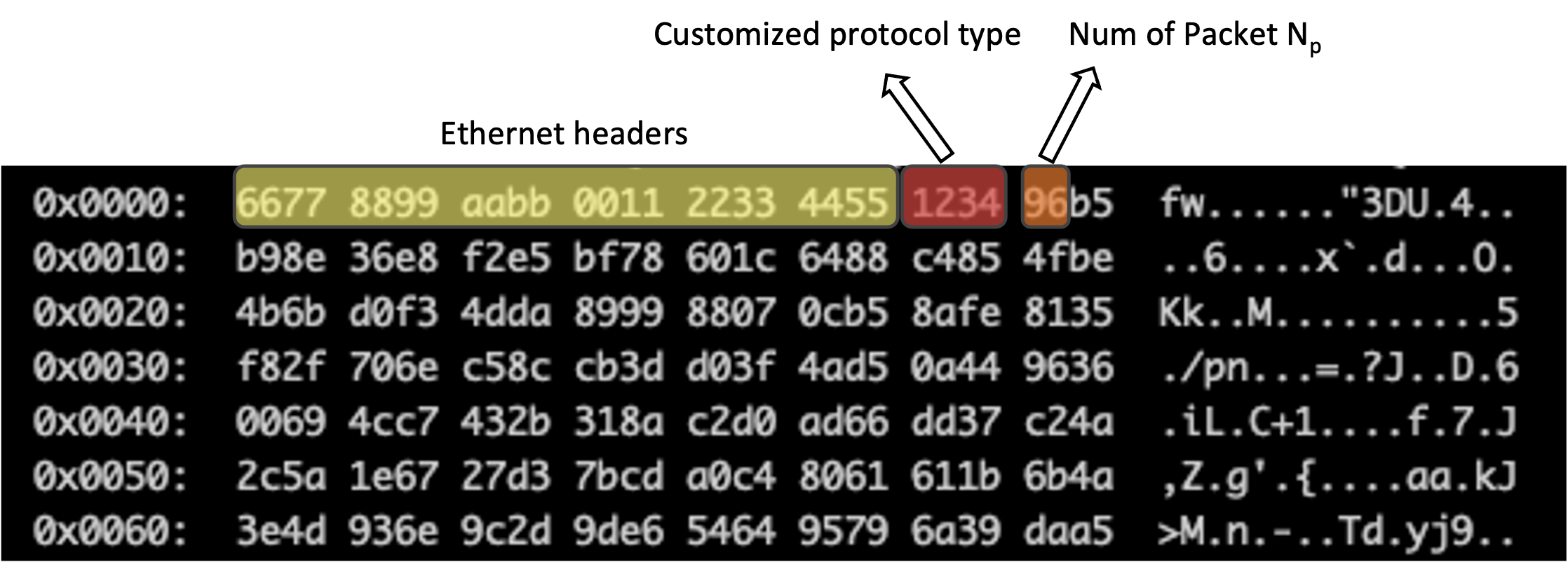}
    \caption{\label{fig:pkt}Received packet example showing the first hundred bytes of the packet.}
    \vspace{-10pt}
  \end{figure}

The packet that summarizes the information contains an Ethernet header with a customized protocol as shown in Fig.~\ref{fig:pkt}. With the help of the P4 deparser, we can easily define and reconstruct any customized protocol. The byte following the header contains the number of source-destination pairs included in the payload, i.e.~$N_p$. In our implementation, we set  $N_p$ to 150 in order to transfer as much information as possible. Then we pack the source and destination IP pairs into the payload. For 150 packets, there will be a 1200-byte payload.

\section{Future Improvements}
\label{sec:discuss}

Our approach, as described in this paper, tackles the first two steps of the Anonymized Network Traffic Graph Challenge, namely extracting source and destination IP addresses at line speed.  Our solution makes use of a network connected FPGA to receive and process packets and transmit the extracted information to the host.  

For ease of implementation, we gathered source and destination IPs, stored them in memory on the FPGA, and used network packets to transfer this information to the host.  We accomplish this by replacing one out of every 150 packets with the collected information. This solution did not require the addition of any additional mechanism to implement host-to-FPGA communication.  It also results in one out of every 150 packets of original data being dropped and not forwarded to the host to accommodate the IP addresses.  

In the future, we plan to investigate different mechanisms for transferring the information to the host machine, independent of the packet data bus.  
For example, it is possible to establish an independent data bus for the desired information through the QDMA, which will require modifications on both the FPGA and host sides.

Second, due to the limitations of the on-chip BRAM, we cannot store a significant amount of data on the FPGA. Our design assumes that steps 3 and 4 in Fig.~\ref{fig:steps} will be executed on the host computer and only IP address extraction is executed on the FPGA. However, the FPGA is equipped with an 8GB high-bandwidth memory (HBM) offering 460GBps bandwidth as well as an additional 32GB DDR4 memory with a lower bandwidth of around 30GBps. A potential improvement to the design is to utilize this memory to construct the data tables directly on the FPGA.

By incorporating these changes,  a comprehensive solution for the Anonymized Network Sensing Challenge can be implemented based on the initial design presented in this paper.

\section{Conclusion}
\label{sec:concl}

This paper proposes an FPGA-based hardware solution for the first few steps of the Anonymized Network Traffic Graph Challenge. Our solution leverages in-network processing to extract and construct anonymized network traffic data structures, demonstrating high throughput that maximizes the potential of an FPGA-based NIC. The design is implemented through a P4-based framework deployed on the Open Cloud Testbed (OCT). \texttt{extern} functions in P4 and High Level Synthesis (HLS) are used to implement portions of the design that require state, as P4 is a stateless language. This paper describes the first example of the combination of using P4 and High-level Synthesis in developing network functions on the latest AMD FPGAs.  

In the future we plan to investigate implementing more of the Anonymized Network Traffic Graph Challenge by improving the communication of information between FPGA and local host computer, as well as implementing more of the graph challenge on the FPGA itself.

\bibliographystyle{plain}
\bibliography{paper}

\end{document}